\documentclass[reprint,showkeys,amsmath,amssymb,aps]{revtex4-2}

\usepackage{graphicx}
\usepackage{dcolumn}
\usepackage{subfigure}
\usepackage{bm}

\usepackage{amsthm}
\newtheorem{lemma}{Lemma}
\usepackage{color}
\definecolor{blue}{rgb}{0.3,0.3,0.9}
\definecolor{green}{rgb}{0.2,0.7,0.2}

\begin{document}

\preprint{APS/123-QED}

\title{Hypercontractivity and factorial moment scaling in the symmetry broken phase}

\author{A.~Brofas}
\email{thanosb@phys.uoa.gr}
\affiliation{Department of Physics, University of Athens, Athens GR-15784, Greece}
\author{M. Zampetakis}
\email{manolis.zampetakis@yale.edu}
\affiliation{Computer Science Department, Yale University, New Haven US-06511, USA}
\affiliation{Archimedes Research Unit, Athens GR-15125, Greece}
\author{F.~K.~Diakonos}
\email{fdiakono@phys.uoa.gr}
\affiliation{Department of Physics, University of Athens, Athens GR-15784, Greece}
\date{\today}


\begin{abstract}

The search for remnants of the QCD chiral critical point is a central objective of current and future high-energy ion collision experiments. Previous studies suggest that a scaling law relating higher-order factorial moments of hadron multiplicity fluctuations to the second factorial moment could serve as a tool for detecting the QCD critical point. However, we demonstrate that this scaling law is not unique to critical phenomena. Instead, it emerges as a general property of distributions by extending the concept of hypercontractivity—originally applied to ordinary moments—to factorial moments. We present examples of distribution classes that exhibit the same higher-order factorial moment scaling as multiplicity fluctuations in the symmetry-broken phase. This insight allows us to explain the recent intermittency analysis results from the STAR experiment at RHIC \cite{STAR_Interm}, where no indication of criticality was observed.

\end{abstract}


\maketitle



The detection of the QCD critical endpoint, remnant of the restoration of chiral symmetry at high temperature and finite baryon density, is the ultimate goal of the recent ion collision experiments STAR at RHIC (BNL) \cite{STAR} and NA61/SHINE at SPS (CERN) \cite{NA61}. Many theoretical efforts during the last decades target to the conception of suitable observables carrying information related to the proximity to the critical point, easily accessible to the experimental analysis \cite{Observables}. In all these efforts significant constrains are provided by the experimental set-up used for the critical point detection. Namely, traces of criticality should be searched in the particle spectra detected from the decay of the state created in collisions of ion beams at high energies. Since in all these experiments the detected particles are hadrons, the chiral symmetry restoration can only occur as a transient effect. Thus, the relevant observables should be based exclusively on the recorded hadronic momenta and their fluctuations. In addition, the hadronic species which are of interest in this search should be associated with the order parameter characterizing the chiral transition \cite{Antoniou2006}. Among the most promising tools in this attempt is the factorial moment analysis of particle number distribution in small cells of momentum space. Proposed in \cite{Bialas} in the context of particle physics, the factorial moment analysis is sensitive to critical fluctuations exhibiting the phenomenon of intermittency \cite{Intermittency} which resembles critical opalescence in strongly interacting matter \cite{Antoniou2006}. In this context the term intermittency refers to the scaling law
\begin{equation}
F_q(M) \sim M^{(q-1) d_F},~~~M \gg 1
\label{eq:scalfm}
\end{equation}
for the suitable normalized factorial moments $F_q(M)$ defined by:
\begin{equation}
F_q(M) =\frac{\langle n(n-1) \dots (n-q+1) \rangle}{\langle n \rangle^q} 
\label{eq:fmom}
\end{equation}
In Eqs.~(\ref{eq:scalfm},\ref{eq:fmom}) $M$ is the number of cells in momentum space, $q$ is the order of the factorial moment and $d_F$ is the fractal dimension related to the geometry of the clusters of the produced particles in momentum space. In Eq.~(\ref{eq:fmom}) the nominator estimates the mean number of $q$-plets of particles within a cell of very small volume $\propto {1 \over M}$ and averaging is over cells and events. 

The scaling law in Eq.~(\ref{eq:scalfm}) implies another scaling connecting the $q$-th factorial moment with the second one:
\begin{equation}
F_q(M) \sim F_2(M)^{(q-1)}
\label{eq:fq2}
\end{equation}
Thus, when the phenomenon of intermittency (Eq.~\ref{eq:scalfm}) is present and the particle momenta recorded in the detector in an ion collision experiment originate from a source at a critical state, one expects Eq.~(\ref{eq:fq2}) to hold. In addition the fractal dimension $d_F$ in Eq.~(\ref{eq:scalfm}) describing the geometry in momentum space is determined by the isothermal critical exponent $\delta$ of the universality class of the transition \cite{Antoniou2019}. However, the appearance of such an ideal, monofractal behaviour in the factorial moment analysis of particle momenta observed in ion collisions, is questionable for two main reasons: (i) the detected particles are necessarily in the hadronic phase with broken chiral symmetry and (ii) experimental resolution and finite statistics hold up from approaching the limit $M \to \infty$ where the intermittency effect is expected to occur. As an alternative to this path for the detection of the QCD critical endpoint through intermittency analysis, in \cite{Hwa1992} was proposed to search for a scaling of the form in Eq.~(\ref{eq:fq2}) in the symmetry broken phase. Using a Ginzburg-Landau free energy for the thermodynamic description of order parameter density fluctuations, it was shown that a similar scaling:
\begin{equation}
F_q(M) \sim F_2(M)^{\beta_q}~~~~,~~~~\beta_q=(q-1)^{\nu},~\nu \approx 1.304
\label{eq:fq2hwa}
\end{equation}
is valid in the symmetry broken (hadronic) phase. Thus, in \cite{Hwa1992} it was proposed to search for the scaling in Eq.~(\ref{eq:fq2hwa}) as a signature for the proximity to the critical point from the hadronic phase. The main task in this case is to measure the exponent $\nu$ expecting to find a value close to $1.3$. This is in contrast to the behaviour in Eq.~(\ref{eq:fq2}) which suggests a value $\nu_c=1$ assuming that the particle source at the critical point possesses a monofractal structure.

This alternative way to search for the critical point was recently adopted in an analysis of the STAR Collaboration at RHIC using momenta of charged particles produced in Au+Au collisions at different beam energies and centralities \cite{STAR_Interm}. As a surprising result they found a scaling behaviour of the form in Eq.~(\ref{eq:fq2hwa}) for all considered systems with a $\nu$-value significantly smaller than $1$, which varies slightly with beam energy and centrality. This puzzling behaviour was not explained in \cite{STAR_Interm} and certainly initiates some queries concerning its physical origin. A first remark in this direction is that the analysis in \cite{STAR_Interm} was performed for all charged particles and therefore the link to order parameter fluctuations is fuzzy. Furthermore, since the observed power-law behaviour appeared in all analysed processes, independently of the energy and centrality of the ion collisions, it can hardly be a benchmark of criticality. Finally, due to finite statistics the higher moment calculations in \cite{STAR_Interm} have been performed up to $q = 6$. Thus, it is not clear if the observed scaling applies also for $q > 6$.

The aim of the present letter is to shed some light to the results of the analysis presented in \cite{STAR_Interm} providing some clarifications as well as some supporting analytical results. Our main claim here is that the scaling behaviour in Eq.~(\ref{eq:fq2hwa}) cannot be associated with the proximity to a critical point. To justify our claim we present the following two arguments:
\begin{enumerate}
  \item Most of the well-studied probability distributions \underline{cannot} have a scaling behavior that grows faster than Eqs.~(\ref{eq:scalfm},\ref{eq:fq2}) as $q \to \infty$. The proof of this claim relies on a fundamental phenomenon in non-asymptotic probability theory called \textit{concentration of measure} \cite{Boucheron2013}, which implies that the average of bounded independent random variables is tightly concentrated around its expectation. This concentration phenomenon is known to be tightly related with \textit{hypercontractivity}, a property of the moments of a distribution, initially discovered in quantum field theory \cite{Nelson1973}.
  \item For small values of $q$, e.g., $q \le q_{max}$ with $q_{max}\lesssim 10$, the behavior suggested by Eq.~(\ref{eq:fq2hwa}) can be observed from a very broad class of distributions, even those that eventually, as $q \to \infty$ will follow the scaling behavior of Eq.~(\ref{eq:scalfm},\ref{eq:fq2}).
\end{enumerate}
To justify our claim number one, we first make the simple observation that:
\begin{equation}
F_q(M) < \frac{\langle n^q \rangle}{\langle n \rangle^q}
\label{eq:moments}
\end{equation}
Next, we use the notion of \textit{sub-Weibull distributions} from \cite{Vladimirova2020} which generalizes the classical concepts of sub-Gaussian and sub-exponential distributions. From Theorem 1 of \cite{Vladimirova2020} we have that for a constant $L$
\begin{lemma}
  If the random variable $n$ follows a sub-Weibull distribution with parameter $\theta~>~0$ then $\langle n^q \rangle \le L^{\theta \cdot q \cdot \ln(q)}$, where $L$ and $\theta$ do not depend on $q$.
\end{lemma}
Combining this lemma with Eq.~(\ref{eq:moments}) we get that if $n$ follows a sub-Weibull distribution with parameter $\theta$ then
\begin{equation}
F_q(M) < \frac{L^{\theta \cdot q \cdot \ln(q)}}{\langle n \rangle^q}
\label{eq:fqBound}
\end{equation}
Most of the known probability distributions, e.g., Gaussian, exponential, Poisson, and more, are sub-Weibull with parameter $\theta < 3$. Hence Eq.~(\ref{eq:fqBound}) suggests that it is impossible for any of these distributions to satisfy a scaling behavior similar to Eq.~(\ref{eq:fq2hwa}). This is because the function $q \cdot \ln(q)$ grows much slower than $q^{1.3}$ and hence for large enough $q$, Eq.~(\ref{eq:fq2hwa}) will be violated for almost all the common distributions.

To justify our second claim we employ the Negative Binomial (NB) and the discrete Weibull (dW) distributions which are both sub-Weibull with a small parameter $\theta$ and both have been used to describe conventional multiplicity fluctuations in $pp$ and $p\bar{p}$ collisions at SPS and LHC energies \cite{SPS,LHC}. We also provide a more generic argument to justify our second claim in the Supplemental Material. Starting our investigations with the NB distribution we obtain the associated $q$-th order factorial moments as:
\begin{equation}
F_{q,NB}=\frac{\Gamma[r+q]}{r^q \Gamma[r]}
\label{eq:fmnbd}
\end{equation}
with $r=\langle n \rangle \left({1 \over p}-1\right)$ and $p \in [0,1]$ the parameters of the distribution. The index $NB$ is used in the following to indicate the NB distribution. Based on Eq.~(\ref{eq:fmnbd}) one can determine the quantity 
\begin{equation}
R_{NB}(q,r)=\mathrm{ln}\left(\frac{\mathrm{ln}(F_{q,NB}(r))}{\mathrm{ln}(F_{2,NB}(r))}\right)
\label{eq:RNB}
\end{equation}
which depends exclusively on $q$ and $r$. In Fig.~1(a) we present the plot of this function using as independent variable the quantity $x=\mathrm{ln}(q-1)$ and setting $r=0.344$. The plot displays the $x$-range $[\mathrm{ln}2,\mathrm{ln}9]$ which corresponds to $3 \leq q \leq 10$ with $q \in \mathbb{N}$. The dashed red line presents the result of a linear fit to this function. It is evident from the plot that the linear behaviour describes very well the variation of $R_{NB}$ with $x$. Thus, both relations $F_{q,NB}=\left(F_{2,NB}\right)^{\beta_{q,NB}}$ and $\beta_{q,NB}=(q-1)^{\nu_{NB}}$ are very well satisfied. The slope of the red dashed  line in Fig.~1(b) determines the index $\nu_{NB}$ in analogy to the definition in Eq.~(\ref{eq:fq2hwa}). It is found $\nu_{NB} \approx 1.304$, a value which is very close to the $\nu$-value obtained in \cite{Hwa1992} using a free-energy approach and introduced there as a signature for the proximity to critical behaviour. For the case of the NB distribution, the linear dependence of $R$ on $x$, appearing in Fig.~1(a) for $r=0.344$, turns out to be a generic property of its factorial moments for any value of the parameter $r$. Thus, one can apply a similar procedure as that followed to obtain Fig.~1(a) and extract through linear regression the slope $\nu_{NB}$ for any value of $r$. As a result of this analysis we display in Fig.~1(b) the values $\nu_{NB}$ for different $r$ with blue stars. Actually, one can also obtain an approximate analytical expression for $\nu_{NB}(r)$ by examining the power law at an intermediate value of $q$ with respect to its range.
\begin{equation}
\nu_{NB,an}(r)=\frac{3(\Psi(r+4)-\mathrm{ln}(r))}{\mathrm{ln}(r+3) + \mathrm{ln}(r+2) + \mathrm{ln}(r+1) - 3 \mathrm{ln}(r)}
\label{eq:nunbd}
\end{equation}
where $\Psi$ is the digamma function. This analytical formula is shown by the dashed red line in Fig.~1(b) and is in good agreement with the fitting results. According to the plot in Fig.~1(b) for $r \approx 0.3$ the corresponding $\nu$-values approach arbitrarily close the value found in \cite{Hwa1992}. Furthermore, there is an infinite family of negative binomial distributions possessing this property. They are obtained by varying the parameter $p$ for a fixed $r$ value. Thus, our initial claim is validated by the NB distribution which provides a statistical model without any connection to critical behaviour.


\begin{figure}[htpb]
    \subfigure[$R_{NB}$ vs $x$.]{{\includegraphics[width = \columnwidth]{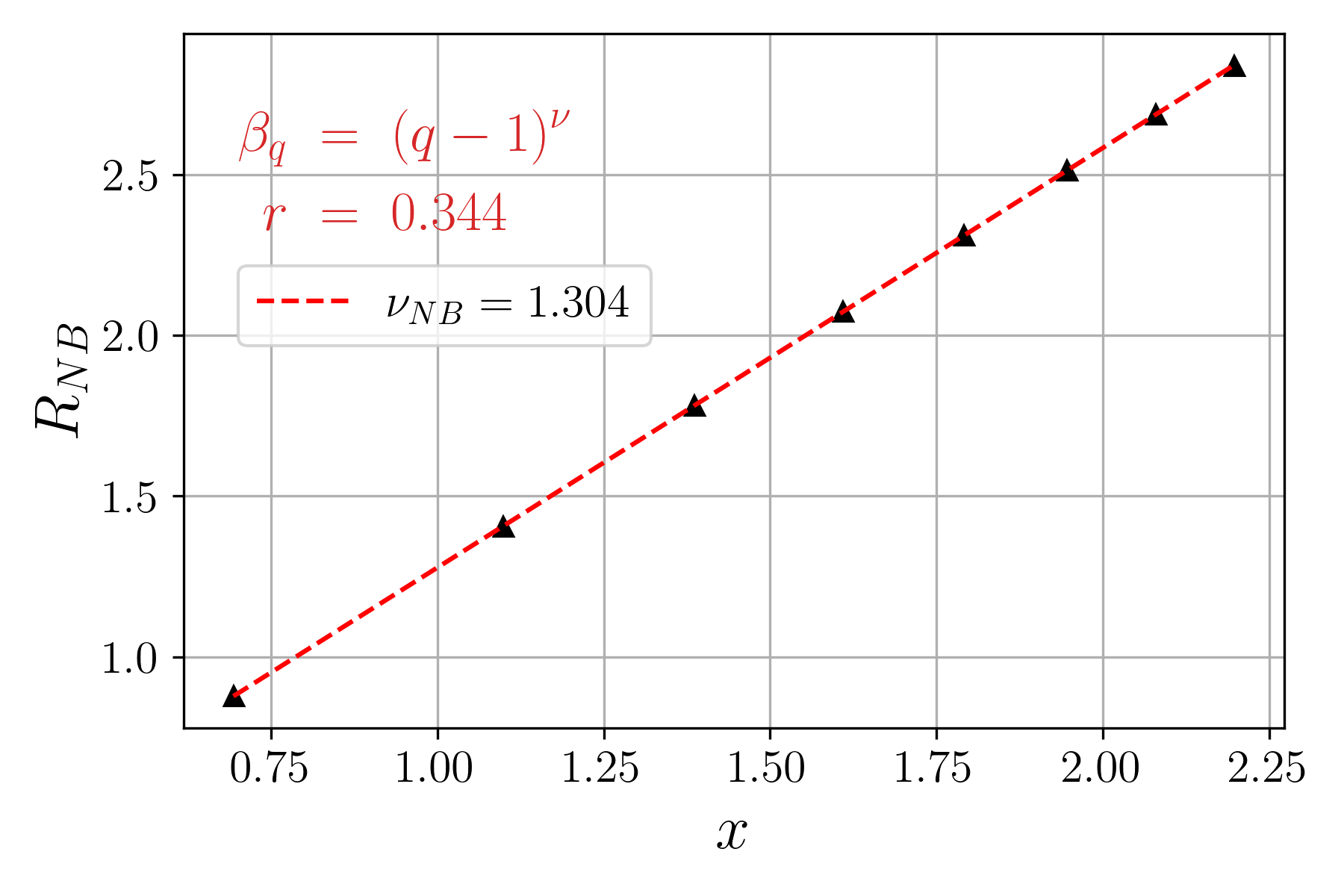}}}\label{NBGa}\hfill
    \subfigure[$\nu_{NB}$ vs $r$ for various values of $r$.]{\includegraphics[width = \columnwidth]{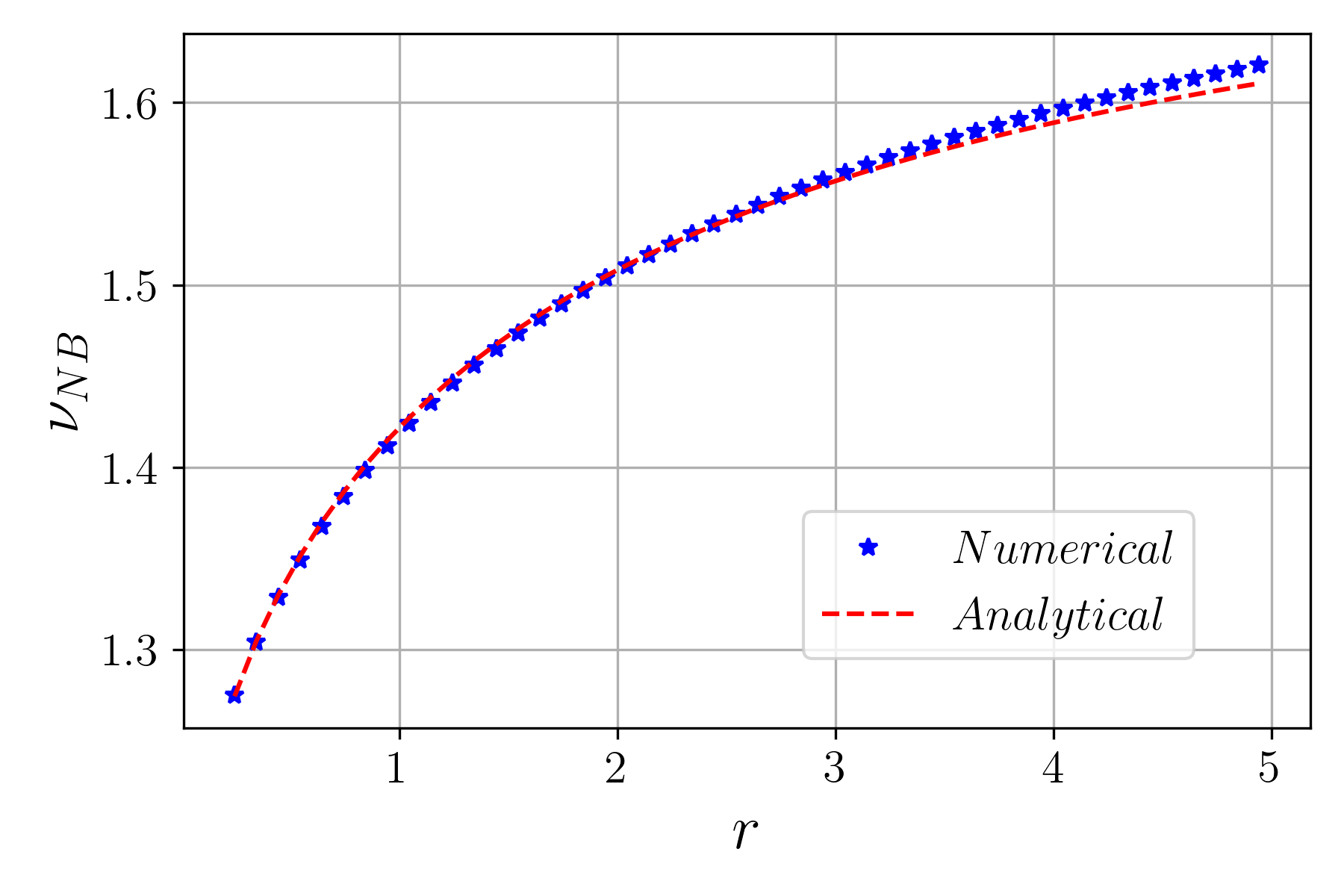}}\label{NBGb}

    \caption[NBD Figures]{(a) The quantity $R_{NB}(q,0.344)$ for the NB distribution as a function of $x=\ln(q-1)$. The dashed line is the result of a linear fit leading to a slope ($\nu_{NB}$) approximately equal to $1.304$. In (b) is displayed the value of $\nu_{NB}$ for different values of $r$ in NB distribution. Notice that $\nu_{NB}(r)$ does not depend on the value of the second parameter $p$ of the NB distribution.} 
    
\end{figure}
As a next step we have calculated the quantity 
\begin{equation}
R_{dW}(q,\lambda,k)=\mathrm{ln}\left(\frac{\mathrm{ln}(F_{q,dW}(\lambda,k))}{\mathrm{ln}(F_{2,dW}(\lambda,k))}\right)
\label{eq:RW}
\end{equation}
for the dW distribution with shape parameter $k$ and scale parameter $\lambda$, using the analytical relations given in \cite{Dash2017}. The power-law dependence of the higher moments on the second order one is well satisfied also for the case of the dW distribution. For $\lambda=1$ and small values of $k$ it turns out that the exponent $\nu_{dW}$ can be estimated analytically leading to the expression:
\begin{equation}
\nu_{dW}=1.295 + 0.053 k - 0.076 k \ln(k) + 0.020 k^2 \ln(k)^2 
\label{eq:nuW}
\end{equation}
We observe that for $k \approx 0.03$ the exponent $\nu_{dW}$ approaches the value $1.304$. In fact there is an infinite set of $(\lambda,k)$ pairs leading to $n_{dW} \approx 1.304$. An argument supporting this statement can be obtained through expansion --in the parametric region $k \ll 1$-- of the factorial moment expressions for the dW distribution in \cite{Dash2017}. In this regime the $\lambda$-dependence drops out to leading order.

Based on the results presented in Figs.~1(a,b) and in Eq.~(\ref{eq:nuW}) we conclude that interpreting the scaling behaviour in Eq.~(\ref{eq:fq2hwa}) and the resulting value of the exponent $\nu$ as a signature of criticality is misleading. As already mentioned, even within NB and dW families, there are infinite many combinations of parameter values leading to the same exponent $\nu \approx 1.304$ when the analysis is performed in a restricted $q$-range. In fact, similar arguments can be transferred also to the case of distributions generated by a free energy functional of the form used in \cite{Hwa1992}. Some details and analytical results for this special case can be found in the Supplemental Material. 

Let us now focus on the results found in the analysis of \cite{STAR_Interm}. There are some methodological issues  in this work which have to stressed out before presenting our interpretation. Firstly, a strong assumption made, is that the scaling law in Eq.~(\ref{eq:fq2hwa}) holds also for differences of factorial moments of multiplicity distributions which possess the same single particle projection. To our knowledge, there is no theoretical work supporting the validity of this property. Secondly, the subtraction of the background correlations at all orders is made through the oversimplified relation:
\begin{equation}
\Delta F_q(M)=F_{q,d}(M)-F_{q,m}(M)
\label{eq:dfq}
\end{equation}
which, as shown in \cite{NA49prot}, is not exactly valid even for $q=2$. In Eq.~(\ref{eq:dfq}) the subscripts "d" and "m" are used for "data" and "mixed events" respectively. Despite the aforementioned ambiguities, the analysis in \cite{STAR_Interm} leads to a surprising result, finding indeed a power-law behaviour of the form: 
\begin{equation}
\Delta F_q(M) \sim \Delta F_2(M)^{\bar{\beta}_q}~~~~,~~~~\bar{\beta}_q=(q-1)^{\bar{\nu}_q}
\label{eq:dfqpl}
\end{equation}
with a $\bar{\nu}_q$-value significantly less than one, for all analysed data sets and for $q \le 6$. We claim that this, rather unexpected, result is related to the use of factorial moment differences in Eq.~(\ref{eq:dfqpl}) instead of factorial moments of a single distribution as in Eq.~(\ref{eq:fq2hwa}). 

As a next step we attempt to clarify how a power-law relation like in Eq.~(\ref{eq:dfqpl}) with an exponent ${\bar{\nu}}_q < 1$ can occur. To this end we assume that the following relations hold (with $A=d, m$):
\begin{equation}
F_{q,A} = a(q) F_{2,A}^{\beta_q}~;~\Delta F_{q}=\bar{a}(q) \Delta F_{2}^{\bar{\beta}_q}
\label{eq:assumptions}
\end{equation}
\noindent In Eq.~(\ref{eq:assumptions}), to simplify our calculations, we consider the amplitudes $a(q)$ and exponents $\beta_q$ for the data and the corresponding mixed events to be the same. In fact this assumption turns out to be true for the Monte-Carlo generated events we will use in the subsequent analysis. Possibly, this is a more general property related to the fact that data and mixed events share the same one-particle density. The  set of relations in Eqs.~(\ref{eq:assumptions}) leads to the following expression for ${\bar{\nu}}_q$:
\begin{equation}
{\bar{\nu}}_q = \frac{1}{\mathrm{ln}(q-1)} \mathrm{ln} \left[ 1+ \underbrace{\frac{1}{\mathrm{ln}\Delta F_2}  \mathrm{ln} \left( \frac{\beta_q F_{2,m}^{\beta_q - 1}}{d(q)}\right)}_{S} \right] 
\label{eq:nubar}
\end{equation}
where $d(q) = \bar{a}(q)/a(q)$. Based on Eq.~(\ref{eq:nubar}) we obtain the condition:
\begin{equation}
  S < 1 ~\Rightarrow a(q) ~\beta_q  F_{2,m}^{\beta_q - 1} < \bar{a}(q) \Delta F_2
   \label{eq:condition}
\end{equation}
for ${\bar{\nu}}_q$ to be less than 1. It is important to notice that in inequality (\ref{eq:condition}) the appearance of $\bar{a}(q)$ --being much greater than $a(q)$-- is crucial, since it allows $\Delta F_q$ to increase significantly with $q$ while keeping ${\bar{\beta}}_q > 0$. Details concerning the derivation of Eq. (\ref{eq:nubar}) are given in the Supplemental Material.

To further explore this scenario we use simulated data obtained through a very simple algorithm. We generate two data sets of 20000 events containing transverse momenta of particles. The particle multiplicity per event is taken to follow a Poisson distribution with mean value $\langle n_p \rangle=10000$. The transverse momenta of the particles in an event are Gaussian random variables with zero mean value and variance $\sigma_p$. In the first data set we used $\sigma_p^{(1)}=0.83$ GeV while in the second data set $\sigma_p^{(2)}=0.84$ GeV. We calculate the factorial moments $F_q^{(1)}(M)$, $F_q^{(2)}(M)$ for each data set in the transverse momentum space region $[-2,2]~\mathrm{GeV} \otimes [-2,2]~\mathrm{GeV}$ using $M$ intervals per direction with $1 \leq M \leq 120$. Then we determine the associated "correlators" defined by the differences $\Delta F_{q}(M)=F_q^{(1)}(M)-F_q^{(2)}(M)$ and we check the validity of a power-low behaviour:
\begin{equation}
\Delta F_q(M)=\bar{a}(q) \Delta F_2(M)^{\bar{\beta}_q}~~\mathrm{with}~~\bar{\beta}_q=(q-1)^{{\bar{\nu}}_q}
\label{eq:dfq12}
\end{equation}

\begin{figure*}[htpb]
\includegraphics[width = 1.7\columnwidth]{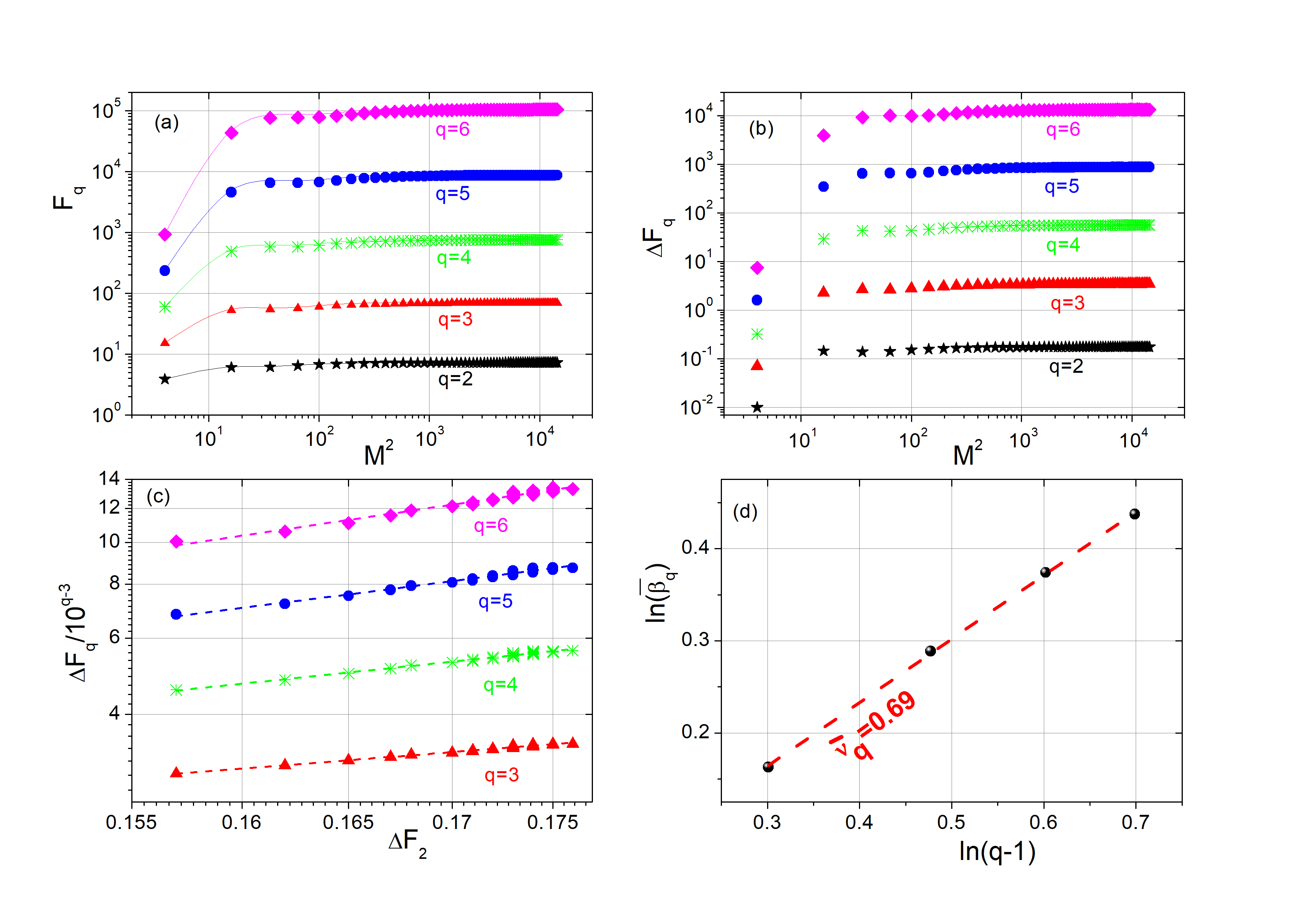}
\caption[Simulation results]{Factorial moment analysis of two data sets, each consisting of $20000$ events containing normally distributed transverse momenta of particles around $(0,0)$ (see description in text). In both data sets the multiplicity per event follows a Poisson distribution with mean $10^4$. The transverse momentum normal distribution in data set $1$ has variance $\sigma_p^{(1)}=0.83$ GeV while in data set $2$ $\sigma_p^{(2)}=0.84$ GeV. In (a) the factorial moments up to order $q=6$ of the two data sets $F_q^{(1)}(M)$ (symbols) and $F_q^{(2)}(M)$ (lines), as a function of the number of cells in transverse momentum space, are shown. In (b) is plotted the "correlator" $\Delta F_{q}(M)=F_q^{(1)}(M)-F_q^{(2)}(M)$ as a function of $M$ for the same range of $q$ values. In (c) is displayed the plot of $\Delta F_{q}(M)=F_q$ for $6 \geq q \geq 3$ versus $\Delta F_{2}(M)$. Finally, in (d) we show $\ln(\bar{\beta}_q)$ versus $\ln(q-1)$ calculated from (c). The dashed line is a linear fit leading to $\bar{\nu}_q \approx 0.69$.}
\label{Simulations}
\end{figure*}

 The results of our calculations are shown in Fig.~3(a-d). In Fig.~3(a), we plot the factorial moments $F_q^{(1)}(M)$ (symbols) and $F_q^{(2)}(M)$ (lines) versus $M$ for both data sets. For $M \gg 1$, the moments saturate to a constant, as no intermittency effect is expected. Fig.~3(b) displays the correlators $\Delta F_q(M)$ versus $M$, again showing no intermittency. However, Fig.~3(c) confirms the power-law from Eq.~(\ref{eq:dfq12}) by plotting $\Delta F_q(M)$ ($q \geq 3$) against $\Delta F_2(M)$ on a double logarithmic scale. Higher moments ($q > 2$) are scaled by a constant factor $10^{q-3}$, showing the power-law increase up to $q=6$ without affecting the slope. Fig.~3(d) shows $\ln(\bar{\beta}_q)$, obtained from the slopes in Fig.~3(c), as a function of $\ln(q-1)$, revealing a linear behavior with $\bar{\nu}_q \approx 0.69$. We tested the validity of the inequality (\ref{eq:condition}) in a large set of pairs of simulated data and mixed events. We found that it is always valid for $q > 3$. In some cases it turned out to be slightly violated for $q=3$. 
 This suggests that the condition (\ref{eq:condition}) is sufficient but not necessary for $\bar{\nu}_q < 1$. To test generality of our claims, we ran additional simulations using a modified cluster model \cite{Seibert1990}, observing similar behavior for $\Delta F_q(M)$ versus $\Delta F_2(M)$. Details are provided in the Supplemental Material.


In summary, we have shown that the power-law behavior in  Eq.~(\ref{eq:fq2hwa}) for $q \le q_{max}$, with $q_{max} \lesssim 10$, is a general property of distributions and cannot be used as an indicator of proximity to a critical point. Scaling behavior for $q \to \infty$, which is experimentally inaccessible, is ruled out by hypercontractivity. Thus, the factorial moment analysis in \cite{STAR_Interm} cannot signal critical proximity. The results in \cite{STAR_Interm} are simply explained by the scaling of factorial moment differences, where $\nu < 1$ naturally arises when certain conditions are met. We derived a sufficient condition for this property and verified our argument using a simple simulation of particle transverse momenta in ion collisions. While the exact value of $\nu$ may provide insights into higher-order statistical properties of transverse momenta in ion collisions, it does not indicate critical behavior.


\begin{thebibliography}{0}%
\makeatletter
\providecommand \@ifxundefined [1]{%
 \@ifx{#1\undefined}
}%
\providecommand \@ifnum [1]{%
 \ifnum #1\expandafter \@firstoftwo
 \else \expandafter \@secondoftwo
 \fi
}%
\providecommand \@ifx [1]{%
 \ifx #1\expandafter \@firstoftwo
 \else \expandafter \@secondoftwo
 \fi
}%
\providecommand \natexlab [1]{#1}%
\providecommand \enquote  [1]{``#1''}%
\providecommand \bibnamefont  [1]{#1}%
\providecommand \bibfnamefont [1]{#1}%
\providecommand \citenamefont [1]{#1}%
\providecommand \href@noop [0]{\@secondoftwo}%
\providecommand \href [0]{\begingroup \@sanitize@url \@href}%
\providecommand \@href[1]{\@@startlink{#1}\@@href}%
\providecommand \@@href[1]{\endgroup#1\@@endlink}%
\providecommand \@sanitize@url [0]{\catcode `\\12\catcode `\$12\catcode `\&12\catcode `\#12\catcode `\^12\catcode `\_12\catcode `\%12\relax}%
\providecommand \@@startlink[1]{}%
\providecommand \@@endlink[0]{}%
\providecommand \url  [0]{\begingroup\@sanitize@url \@url }%
\providecommand \@url [1]{\endgroup\@href {#1}{\urlprefix }}%
\providecommand \urlprefix  [0]{URL }%
\providecommand \Eprint [0]{\href }%
\providecommand \doibase [0]{https://doi.org/}%
\providecommand \selectlanguage [0]{\@gobble}%
\providecommand \bibinfo  [0]{\@secondoftwo}%
\providecommand \bibfield  [0]{\@secondoftwo}%
\providecommand \translation [1]{[#1]}%
\providecommand \BibitemOpen [0]{}%
\providecommand \bibitemStop [0]{}%
\providecommand \bibitemNoStop [0]{.\EOS\space}%
\providecommand \EOS [0]{\spacefactor3000\relax}%
\providecommand \BibitemShut  [1]{\csname bibitem#1\endcsname}%
\let\auto@bib@innerbib\@empty
\end{thebibliography}%


\begin{thebibliography}{99}



\bibitem{STAR_Interm} The STAR Collaboration, Phys. Lett. B {\bf 845}, 138165 (2023), arXiv:2301.11062v1 [nucl-ex]; J. Wu for the STAR Collaboration, SciPost Phys. Proc. {\bf 10}, 041 (2022).

\bibitem{STAR} STAR Note 0598: BES-II whitepaper. https://drupal.
star.bnl.gov/STAR/starnotes/public/sn0598.

\bibitem{NA61} M. Mackowiak-Pawlowska (NA61/SHINE), 
Nucl. Phys. A {\bf 1005}, 121753 (2021), arXiv:2002.04847 [nucl-ex].

\bibitem{Observables} A. Pandav, D. Mallick and B. Mohanty, 
Prog. Part. Nucl. Phys. {\bf 125}, 103960 (2022).

\bibitem{Antoniou2006} N.~G. Antoniou, F.~K. Diakonos, A.~S. Kapoyannis and K.~S. Kousouris, Phys. Rev. Lett. {\bf 97}, 032002 (2006).

\bibitem{Bialas} A. Bialas and R. Peschanski, Nucl. Phys. B {\bf 273}, 703 (1986); {\bf 308}, 857 (1988).

\bibitem{Intermittency} H. Satz, Nucl. Phys. B {\bf 326}, 613 (1989); A. Bialas, Nucl. Phys. A {\bf 545}, 285 (1992); N.~G. Antoniou, F.~K. Diakonos, I.~S. Mistakidis and C.~G. Papadopoulos, Phys. Rev. D {\bf 49}, 5789 (1994); N.~G. Antoniou, F.~K. Diakonos, X.~N. Maintas and C.~E. Tsagkarakis, Phys. Rev. D {\bf 97},  034015 (2018).

\bibitem{Antoniou2019} N.~G. Antoniou and F.~K. Diakonos, 
J. Phys. G (Nucl. Part. Phys.) {\bf 46}, 3 (2019).

\bibitem{Hwa1992} R.~C. Hwa and M.~T. Nazirov, 
Phys. Rev. Lett. {\bf 69}, 741 (1992).

\bibitem{SPS} G.~J. Alner {\it et al.} (UA5 Collaboration), Phys. Lett. B {\bf 138}, 304 (1984); Phys. Lett. B {\bf 160}, 199 (1985).

\bibitem{LHC} V. Khachatryan {\it et al.} (CMS Collaboration), J. High Energy Phys. {\bf 01}, 079 (2011); J. Adam {\it et al.} (ALICE Collaboration), Eur. Phys. J. C {\bf 77}, 33 (2017).

\bibitem{Dash2017} A.~K. Pandey, P. Sett and K. Dash, 
Phys. Rev. D {\bf 96}, 074006 (2017).

\bibitem{NA49prot} T. Anticic {\it et al.} NA49 Collaboration, Eur. Phys. J. C {\bf 75}, 587 (2015), arXiv:1208.5292 [nucl-ex].

\bibitem{Seibert1990} D. Seibert, Phys. Rev. D {\bf 41}, 3381 (1990).

\bibitem{Boucheron2013} S. Boucheron, G. Lugosi and P. Massart, {\it Concentration Inequalities: A Nonasymptotic Theory of Independence}, Oxford University Press, 2013.

\bibitem{Nelson1973} E. Nelson, J. Funct. Anal. {\bf 12}, 97 (1973).

\bibitem{Vladimirova2020} M. Vladimirova, S. Girard, H. Nguyen and J. Arbel, Stat. {\bf 9}, e318 (2020).

\end{thebibliography}
\end{document}


\title{Supplemental materials to the manuscript "Hypercontractivity and factorial moment scaling in the symmetry broken phase"}
\maketitle
\begin{center}
by Athanasios Brofas, Manolis Zampetakis and Fotios K. Diakonos
\end{center}

\section{Construction of distributions with prescribed $\nu$ for $q \le q_{max}$}

To justify the scaling relation:
\begin{equation}
F_q(M) \sim F_2(M)^{\beta_q}~~~~,~~~~\beta_q=(q-1)^{\nu}
\label{eq:fq2sc}
\end{equation}
for small values of $q$, e.g., $q = 3, 4, \dots, q_{max}$, all we need is to satisfy for these values of $q$ that 
\begin{equation}
\langle n (n - 1) \cdots (n - q + 1) \rangle \cdot \langle n \rangle^{2 \beta_q} = \langle n (n - 1) \rangle^{\beta_q} \cdot \langle n \rangle^{q}.
\label{eq:factMoments}
\end{equation}
This creates a system of $q_{max}-2$ equations so it should be expected than for any family of distributions with enough parameters there should be a distribution in the family that satisfies the system of Eq.~(\ref{eq:factMoments}). We made that precise with the Negative Binomial and the discrete Weibull families in the main part but there are many more families for which we can satisfy Eq.~(\ref{eq:factMoments}) for $q = 3, 4, \dots, q_{max}$. 

A general way to construct such examples is by taking mixtures of probability distributions with different parameters. A good family of distributions for this task is to consider mixtures of Poisson distributions. Poisson distributions are a good choice for two reasons: First, Poisson distributions are universal approximators of most positive discrete random variables \cite{le1960approximation} the same way that Gaussians are universal approximators of continuous random variables due to the Central Limit Theorem. Second, the $q$th factorial moments of a Poisson distribution with parameter $\lambda$ takes the very simple form of $\lambda^q$. 

We illustrate the effectiveness of the method proposed above via an example. Let's consider $q_{max} = 6$, $\nu = 1.304$ and we want to find a uniform mixture of $\ell$ Poisson distributions that satisfy \eqref{eq:fq2sc}. Let $\lambda_1, \lambda_2, \dots, \lambda_{\ell}$ be the parameters of these Poisson distributions. Then according to \eqref{eq:factMoments}, and adding a constraint that the mean is equal to $1$ to simplify the calculations, we need for this set of $\lambda_i$'s to satisfy the following equations
\begin{align*}
  \langle n \rangle & = 1 \\
  \langle n (n - 1) \rangle & = \mu_2 \\
  \langle n (n - 1) (n - 2) \rangle & = (\mu_2)^{2^{1.304}} \\
  \langle n (n - 1) (n - 2) (n - 3) \rangle & = (\mu_2)^{3^{1.304}} \\
  \langle n (n - 1) (n - 2) (n - 3) (n - 4) \rangle & = (\mu_2)^{4^{1.304}} \\
  \langle n (n - 1) (n - 2) (n - 3) (n - 4) (n - 5) \rangle & = (\mu_2)^{5^{1.304}}
\end{align*}
and because we assumed that $n$ is a mixture of Poisson distributions we have that
\begin{align*}
  \frac{1}{\ell} \sum_{j = 1}^\ell \lambda_j = 1 \qquad &, \qquad \frac{1}{\ell} \sum_{j = 1}^\ell \lambda_j^2 = \mu_2 \\
  \frac{1}{\ell} \sum_{j = 1}^\ell \lambda_j^3 = (\mu_2)^{2^{1.304}} \qquad &, \qquad \frac{1}{\ell} \sum_{j = 1}^\ell \lambda_j^4 = (\mu_2)^{3^{1.304}} \\
  \frac{1}{\ell} \sum_{j = 1}^\ell \lambda_j^5 = (\mu_2)^{4^{1.304}} \qquad &, \qquad
  \frac{1}{\ell} \sum_{j = 1}^\ell \lambda_j^6 = (\mu_2)^{5^{1.304}}.
\end{align*}
If we pick for example $\mu_2 = 1.1$ then a mixture of $\ell = 6$ Poisson distributions suffices to satisfy the above equations with error less than $10^{-4}$ and the solution that we get is
\begin{align*}
\lambda_1 = 0.3568 \qquad &, \qquad \lambda_2 = 0.9048 \qquad , \qquad \lambda_3 = 1.1248 \\
\lambda_4 = 1.1248 \qquad &, \qquad \lambda_5 = 1.1248 \qquad , \qquad \lambda_6 = 1.3635.
\end{align*}


\section{Scaling exponent $\nu_F$ for a generalized Ginzburg-Landau free energy}

We explored the emergence of a scaling law of the form in Eq.~(\ref{eq:fq2sc}) for multiplicity distributions generated through a generalized free energy functional of the Ginzburg-Landau form:
\begin{equation}
\mathcal{F}[\phi]=\int_{\Omega_r} d\vec{r} \left[a(T) \vert \phi(\vec{r}) \vert^2 + b(T) \vert \phi(\vec{r}) \vert^{2 k} \right] 
\label{eq:freeener}
\end{equation}
where $\phi(\vec{r})$ is the order-parameter field determining the distribution of the multiplicity $n$, $\vec{r}$ is the spatial coordinate, $\Omega_r$ is the available volume and $a(T)$, $b(T)$ are coefficients depending on the temperature $T$. Notice that $a(T) < 0$ and $b(T) > 0$.  Using Eq. (\ref{eq:freeener}) we have numerically calculated the function $\nu_F(k)$ for $k \in (1,3]$ -- along the lines given in \cite{Hwa1992} -- and the results are presented in Fig.~1. It is clearly shown that the index $\nu_F$ varies very slowly with $k$. All the calculations have been performed for $q \le 10$. We use $a(T)$, $b(T)$ values as in \cite{Hwa1992}.


\begin{figure}[h!]
\includegraphics[width = 0.55\columnwidth]{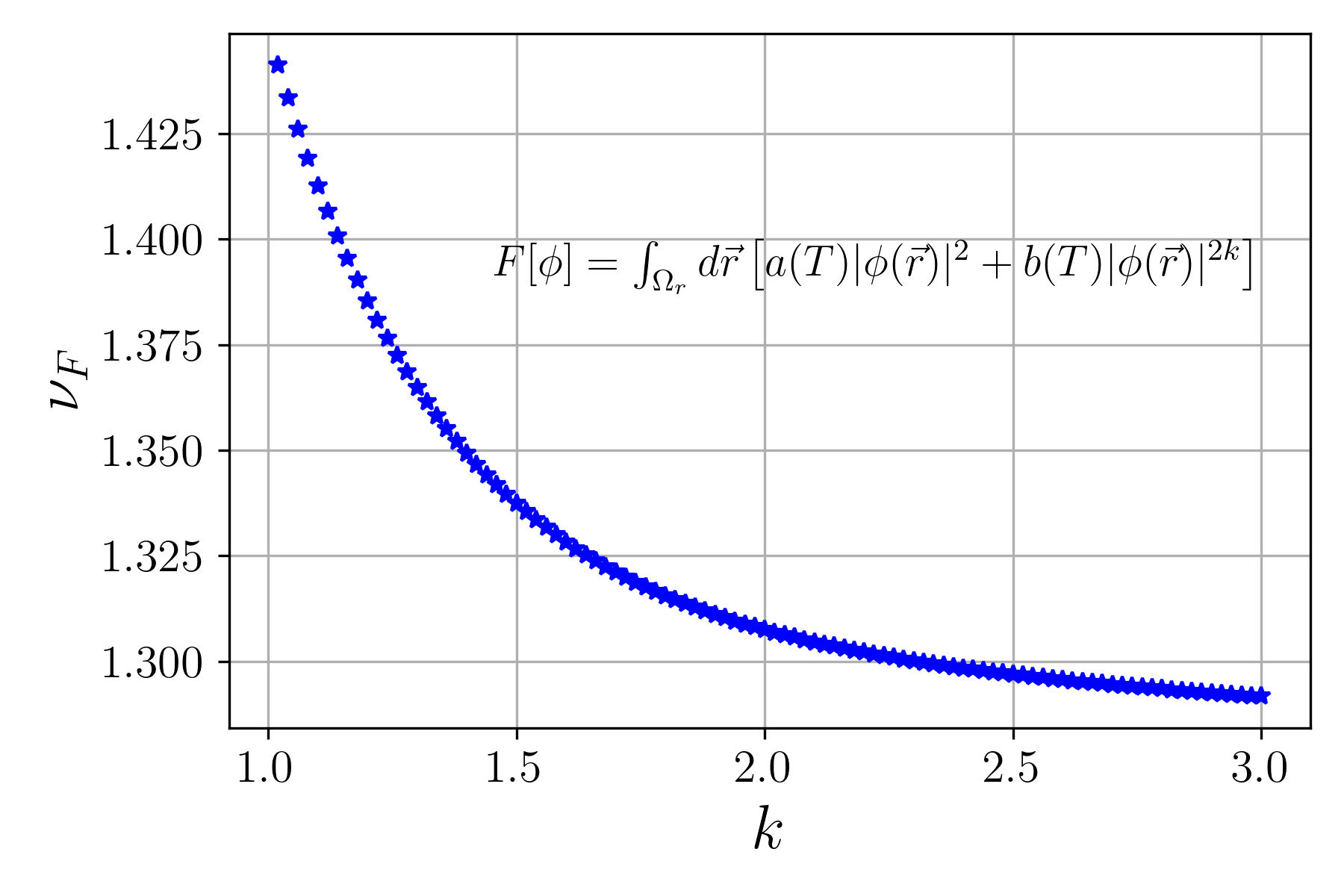}
\caption[Two Term Free Energy]{The exponent $\nu_F$ as a function of $k$ entering in the free energy \ref{eq:freeener}. All the calculations have been performed using the approach described in \cite{Hwa1992}. There is no significant dependence of $\nu_F$ on the $k$-value.}
\label{Exp_two_term}
\end{figure}
Furthermore, following the approach in \cite{Hwa1992} we determined analytically the index $\nu_F$ for the case $k=2$ leading to the expression:
\begin{equation}
\nu_F(2)=\frac{\mathrm{ln}(\pi)+\Psi\left(\frac{3}{2}\right)}{2(\mathrm{ln}(\pi) - \mathrm{ln}(2))}\approx 1.304
\end{equation}
which describes very well the scaling relation in Eq.~(\ref{eq:fq2sc}), valid for $2 < q \le 10$ as mentioned in \cite{Hwa1992}. Notice that for all considered cases $\nu_F \to 1^+$ for $q \to \infty$. This supports further the fact that the scaling behaviour observed in \cite{Hwa1992} can not be related to proximity to a critical point.

\section{"Correlator" scaling}

\noindent 
We determine here a sufficient condition for the validity of the power-law scaling: $\Delta F_q(M)=(\Delta F_2(M))^{\bar{\beta}_q}$ with $\bar{\beta}_q=(q-1)^{\bar{\nu}_q}$ and $\bar{\nu}_q < 1$. Our treatment is based on the following assumptions:
\begin{eqnarray}
    F_{q,d} &= a(q) F_{2,d}^{\beta_q} \nonumber\\
    F_{q,m} &= a(q) F_{2,m}^{\beta_q} \nonumber\\
    d(q) &= \frac{\bar{a}(q)}{a(q)}
    \label{eq:assum}
\end{eqnarray}

\noindent We also assume that $\Delta F_2 \ll 1$ for the entire range of values for the number of cells $M$. It follows that:
\begin{equation*}
    \Delta F_q = \bar{a}(q) \Delta F_2 ^{\bar{\beta}_q} 
     \Rightarrow (F_{2,m}+\Delta F_2)^{\beta_q} - F_{2,m}^{\beta_q} = d(q) \Delta F_2^{\bar{\beta}_q}
\end{equation*} 

\noindent
Then, for $\Delta F_2(M) \ll F_{2,m}(M)$ for all $M$, we obtain:
\begin{equation*}
     F_{2,m}^{\beta_q}\left( 1+ \frac{\Delta
F_2}{F_{2,m}}\right)^{\beta_q} - F_{2,m}^{\beta_q} = d(q) \Delta
F_2^{\bar{\beta}_q} 
     \Rightarrow  F_{2,m}^{\beta_q}\left( 1+ \beta_q \frac{\Delta
F_2}{F_{2,m}}\right) - F_{2,m}^{\beta_q} = d(q) \Delta F_2^{\bar{\beta}_q} 
\end{equation*}

\noindent
which in turn leads to:
\begin{equation*}
     \beta_q F_{2,m}^{\beta_q -1} = d(q) \Delta F_2 ^{\bar{\beta}_q-1} 
     \Rightarrow \bar{\beta}_q - 1 = \frac{\mathrm{ln}\beta_q + (\beta_q - 1)
\mathrm{ln} F_{2.m} - \mathrm{ln}d(q)}{\mathrm{ln} \Delta F_2}
\end{equation*}

\noindent Since $\mathrm{ln}\Delta F_2 < 0$ and $\bar{\beta}_q >1 $, the numerator must be negative. This leads to the constraint:
\begin{equation}
  \frac{\bar{a}(q)}{a(q)} > \beta_q F_{2,m}^{\beta_q -1}
  \label{eq:constr1}
\end{equation}

\noindent Proceeding with the calculations we find:

\begin{equation*}
    \bar{\beta}_q = 1+ \frac{1}{\mathrm{ln}\Delta F_2} \left( \mathrm{ln}\beta_q +
(\beta_q - 1) \mathrm{ln} F_{2.m} - \mathrm{ln}d(q) \right)
\end{equation*}

\begin{equation*}
    \bar{\nu}_q = \frac{1}{\mathrm{ln}(q-1)} \mathrm{ln} \left[ 1+
\underbrace{ \frac{1}{\mathrm{ln}\Delta F_2}  \mathrm{ln} 
\left( \frac{\beta_q F_{2,m}^{\beta_q - 1}}{d(q)}\right) }_{S}
\right]
\end{equation*}

\begin{equation*}
    \bar{\nu}_q = \frac{\mathrm{ln}(1+S)}{\mathrm{ln} (q-1)}
\end{equation*}

\noindent In order for $\bar{\nu}_q$ to be less than $1$ we consider the case $q \geq 3$ and we obtain the condition $S < 1$ leading to:
\begin{equation}
   \beta_q a(q) F_{2,m}^{\beta_q - 1} < \bar{a}(q) \Delta F_2
   \label{eq:constr2}
\end{equation}
This should primarily hold for $q \gtrsim 3$, since as $q$ increases, the right-hand side grows disproportionately compared to the left-hand side. Notice that the condition in Eq.~(\ref{eq:constr2}) is more restrictive than the condition in Eq.~(\ref{eq:constr1}) because $\Delta F_2 \ll 1$. Verifying the justification of the condition (\ref{eq:constr2}) in a Gaussian model simulation, we obtained the following values: $a(q) \approx 1.3 $, $\beta_q \approx 2$, $F_{2,m} \approx 7$, $\bar{a}_q \approx 100$, and $\Delta F_2 \approx 0.2$, yielding $18.2<20$ for $q=3$, which is indeed true.\\


Thus, it is evident that the amplitudes—or more precisely, the offsets determined by $a(q)$ and $\bar{a}(q)$—play a critical role in ensuring that the condition $\bar{\nu}_q < 1$ holds. The values of these quantities can fine-tune $\bar{\nu}_q$ to values less than one, while simultaneously maintaining the condition $\bar{\beta}_q > 1$. This condition is valid by definition, as $\bar{\beta}_q = (q-1)^{\bar{\nu}_q}$ with $q > 2$ and $\bar{\nu}_q > 0$.

\section{Cluster model for hadron transverse momenta}
 
 To further explore the scenarios for the scaling 
 \begin{equation}
\Delta F_q(M) \sim \Delta F_2(M)^{\bar{\beta}_q}~~~~,~~~~\bar{\beta}_q=(q-1)^{{\bar{\nu}}_q}
\label{eq:dfqpow}
\end{equation}
 with $\bar{\nu}_q < 1$ we performed additional simulations with a suitable modification of the cluster model described in \cite{Seibert1990}, adapted to the problem at hand. In this more complicated model the transverse momenta of the particles in one event are normally distributed around a center $(p_{x,c},p_{y,c})$ and with variance $\sigma_p$. The center itself is normally distributed around $(0,0)$ with a variance $\sigma_c$. Similarly to the case of the simple model, which led to the results shown in Fig.~3(a-d), we generated pairs of data sets to calculate $\Delta F_{q}(M)=F_q^{(1)}(M)-F_q^{(2)}(M)$ for each such pair. The data sets consisting the pair, possess the same mean multiplicity, the same $\sigma_c$ and they only differ in $\sigma_p$. For each data set in the pair the multiplicity per event is chosen to be Poisson distributed with a prescribed mean value. We analysed a large number of data set pairs and we observed the following behaviour: the crucial parameter determining the value of $\bar{\nu}_q$ is the difference $\delta \sigma_p=\sigma_p^{(1)}-\sigma_p^{(2)}$. For increasing $\delta \sigma_p$ the value $\bar{\nu}_q$ tends to $\bar{\nu}_q=1$. This behaviour does not depend on the multiplicity per event of the involved data sets provided that the latter is large enough to allow for the calculation of factorial moments of sufficiently high order (up to $q=6$). In our numerical experiments we used mean multiplicities in the range of 250 to 10000 particles per event. In fact increasing the mean multiplicity one can achieve accurate calculation of $\bar{\nu}_q$ (with less than 1\% error) with moderate statistics. Thus, for mean multiplicity 1000 particles per event one needs $10^6$ events to obtain the same accuracy for the $\bar{\nu}_q$ value as in the case with 10000 particles per event and 20000 events. Nevertheless, the obtained $\bar{\nu}_q$ value turns out to be the same (within 1\%) provided that $\delta \sigma_p$ is the same. Unfortunately, if the mean multiplicity per event is 250, which approaches the experimental values in \cite{STAR_Interm}, then the number of events should increase dramatically to achieve the same accuracy. Actually, for mean multiplicity 250 particles per event we performed calculations with $10^7$ events per data set and the fluctuations in the $\bar{\beta}_q$ and $\bar{\nu}_q$ values were still very large. Finally, an interesting property of the simulated events is that for $\delta \sigma_p$ in the range $[0.001,0.01]$ there is practically no variation of the estimated value of $\bar{\nu}_q\approx 0.69$. Thus, it seems that there is a plateau for $\bar{\nu}_q$ in this range of $\delta \sigma_p$ values. It may be that this $\bar{\nu}_q$-plateau value is characteristic for normally distributed transverse momenta of particles. However, such a statement needs further investigation to be set in a more solid basis. If so, it is a challenging question to find the appropriate transverse momentum distribution which could lead to $\nu$-values closer to those measured in STAR \cite{STAR_Interm}.

